\def\bm#1{\mbox{\boldmath $#1$}}
\def\rlx{\relax\leavevmode}
\def\inbar{\vrule height1.5ex width.4pt depth0pt}
\def\IZ{\rlx\hbox{\small \sf Z\kern-.4em Z}}
\def\IR{\rlx\hbox{\rm I\kern-.18em R}}
\def\ID{\rlx\hbox{\rm I\kern-.18em D}}
\def\IC{\rlx\hbox{\,$\inbar\kern-.3em{\rm C}$}}
\def\one{\hbox{{1}\kern-.25em\hbox{l}}}
\def\smallfrac#1#2{\mbox{\small $\frac{#1}{#2}$}}

\documentstyle[twoside]{article}

\catcode`\@=11
\long\def\@makefntext#1{ 
\protect\noindent \hbox to 3.2pt {\hskip-.9pt
$^{{\eightrm\@thefnmark}}$\hfil}#1\hfill} 

\def\thefootnote{\fnsymbol{footnote}}
 \def\@makefnmark{\hbox to 0pt{$^{\@thefnmark}$\hss}}  

\def\ps@myheadings{\let\@mkboth\@gobbletwo
\def\@oddhead{\hbox{} 
\rightmark\hfil\eightrm\thepage}
\def\@oddfoot{}\def\@evenhead{\eightrm\thepage\hfil 
\leftmark\hbox{}}\def\@evenfoot{}
\def\sectionmark##1{}\def\subsectionmark##1{}}

\oddsidemargin=\evensidemargin
\renewcommand{\thefootnote}{\fnsymbol{footnote}}

\newcounter{sectionc}\newcounter{subsectionc}\newcounter{subsubsectionc}
\renewcommand{\section}[1] {\vspace{12pt}\addtocounter{sectionc}{1}
\setcounter{subsectionc}{0}\setcounter{subsubsectionc}{0}\noindent
	{\tenbf\thesectionc. #1}\par\vspace{5pt}}
\renewcommand{\subsection}[1] {\vspace{12pt}\addtocounter{subsectionc}{1}
	\setcounter{subsubsectionc}{0}\noindent
	{\bf\thesectionc.\thesubsectionc. {\kern1pt \bfit #1}}\par\vspace{5pt}}
\renewcommand{\subsubsection}[1]
{\vspace{12pt}\addtocounter{subsubsectionc}{1}
	\noindent{\tenrm\thesectionc.\thesubsectionc.\thesubsubsectionc.
	{\kern1pt \tenit #1}}\par\vspace{5pt}}
\newcommand{\nonumsection}[1] {\vspace{12pt}\noindent{\tenbf #1}
	\par\vspace{5pt}}

\newcounter{appendixc}
\newcounter{subappendixc}[appendixc]
\newcounter{subsubappendixc}[subappendixc]
\renewcommand{\thesubappendixc}{\Alph{appendixc}.\arabic{subappendixc}}
\renewcommand{\thesubsubappendixc}
	{\Alph{appendixc}.\arabic{subappendixc}.\arabic{subsubappendixc}}

\renewcommand{\appendix}[1] {\vspace{12pt}
        \refstepcounter{appendixc}
        \setcounter{figure}{0}
        \setcounter{table}{0}
        \setcounter{lemma}{0}
        \setcounter{theorem}{0}
        \setcounter{corollary}{0}
        \setcounter{definition}{0}
        \setcounter{equation}{0}
        \renewcommand{\thefigure}{\Alph{appendixc}.\arabic{figure}}
        \renewcommand{\thetable}{\Alph{appendixc}.\arabic{table}}
        \renewcommand{\theappendixc}{\Alph{appendixc}}
        \renewcommand{\thelemma}{\Alph{appendixc}.\arabic{lemma}}
        \renewcommand{\thetheorem}{\Alph{appendixc}.\arabic{theorem}}
        \renewcommand{\thedefinition}{\Alph{appendixc}.\arabic{definition}}
        \renewcommand{\thecorollary}{\Alph{appendixc}.\arabic{corollary}}
        \renewcommand{\theequation}{\Alph{appendixc}.\arabic{equation}}
        \noindent{\tenbf Appendix \theappendixc #1}\par\vspace{5pt}}
\newcommand{\subappendix}[1] {\vspace{12pt}
        \refstepcounter{subappendixc}
        \noindent{\bf Appendix \thesubappendixc. {\kern1pt \bfit #1}}
	\par\vspace{5pt}}
\newcommand{\subsubappendix}[1] {\vspace{12pt}
        \refstepcounter{subsubappendixc}
        \noindent{\rm Appendix \thesubsubappendixc. {\kern1pt \tenit #1}}
	\par\vspace{5pt}}

\newcommand{\textlineskip}{\baselineskip=13pt}
\newcommand{\smalllineskip}{\baselineskip=10pt}

\def\eightcirc{
\begin{picture}(0,0)
\put(4.4,1.8){\circle{6.5}}
\end{picture}}
\def\eightcopyright{\eightcirc\kern2.7pt\hbox{\eightrm c}}

\newcommand{\copyrightheading}[1]
	{\vspace*{-2.5cm}\smalllineskip{\flushleft
	{\eightrm International Journal of Modern Physics B, #1}\\
	{\eightrm $\eightcopyright$\, World Scientific Publishing
	 Company}\\
	 }}

\def\abstracts#1#2#3{{
	\centering{\begin{minipage}{4.5in}\baselineskip=10pt\eightrm
	\centerline{ABSTRACT}
	\parindent=0pt #1\par
	\parindent=15pt #2\par
	\parindent=15pt #3
	\end{minipage} }\par}}

\newcommand{\bibit}{\nineit}

\renewenvironment{thebibliography}[1]			
	{\ninerm
	 \baselineskip=11pt				
	 \begin{list}{\arabic{enumi}.}
	{\usecounter{enumi}\setlength{\parsep}{0pt}
	 \setlength{\leftmargin 17pt}{\rightmargin 0pt}	
	 \setlength{\itemsep}{0pt} \settowidth		
	{\labelwidth}{#1.}\sloppy}}{\end{list}}

\newcounter{itemlistc}
\newcounter{romanlistc}
\newcounter{alphlistc}
\newcounter{arabiclistc}

\newcommand{\fcaption}[1]{
        \refstepcounter{figure}
        \setbox\@tempboxa = \hbox{\eightrm Fig.~\thefigure. #1}
        \ifdim \wd\@tempboxa > 5in
           {\begin{center}
        \parbox{5in}{\eightrm \smalllineskip Fig.~\thefigure. #1 }
            \end{center}}
        \else
             {\begin{center}
             {\eightrm Fig.~\thefigure. #1}
              \end{center}}
        \fi}

\newcommand{\tcaption}[1]{
        \refstepcounter{table}
        \setbox\@tempboxa = \hbox{\eightrm Table~\thetable. #1}
        \ifdim \wd\@tempboxa > 5in
           {\begin{center}
        \parbox{5in}{\eightrm\smalllineskip Table~\thetable. #1 }
            \end{center}}
        \else
             {\begin{center}
             {\eightrm Table~\thetable. #1}
              \end{center}}
        \fi}

\def\@citex[#1]#2{\if@filesw\immediate\write\@auxout	
	{\string\citation{#2}}\fi			
\def\@citea{}\@cite{\@for\@citeb:=#2\do			
	{\@citea\def\@citea{,}\@ifundefined		
	{b@\@citeb}{{\bf ?}\@warning
	{Citation `\@citeb' on page \thepage \space undefined}}
	{\csname b@\@citeb\endcsname}}}{#1}}

\newif\if@cghi
\def\cite{\@cghitrue\@ifnextchar [{\@tempswatrue
	\@citex}{\@tempswafalse\@citex[]}}
\def\citelow{\@cghifalse\@ifnextchar [{\@tempswatrue
	\@citex}{\@tempswafalse\@citex[]}}
\def\@cite#1#2{{$\null^{#1}$\if@tempswa\typeout
	{IJCGA warning: optional citation argument
	ignored: `#2'} \fi}}

\def\pmb#1{\setbox0=\hbox{#1}
	\kern-.025em\copy0\kern-\wd0
	\kern.05em\copy0\kern-\wd0
	\kern-.025em\raise.0433em\box0}

\def\fnt#1#2{\footnotetext{\kern-.3em
	{$^{\mbox{\scriptsize #1}}$}{#2}}}

\def\fpage#1{\begingroup
\voffset=.3in
\thispagestyle{empty}\begin{table}[b]\centerline{\footnotesize #1}
	\end{table}\endgroup}

\def\runninghead#1#2{\pagestyle{myheadings}
\markboth{{\eightit{\quad #1}}\hfill}{\hfill{\eightit{#2\quad}}}}

\font\tenbf=cmbx10
\font\tenit=cmti10
\font\tenit=cmti10
\font\bfit=cmbxti10 at 10pt
 1
 1
 1

\font\ninerm=cmr9
\font\nineit=cmti9

\font\eightrm=cmr8
\font\eightit=cmti8

\def\qed{\hbox{${\vcenter{\vbox{                          
   \hrule height 0.4pt\hbox{\vrule width 0.4pt height 6pt
   \kern5pt\vrule width 0.4pt}\hrule height 0.4pt}}}$}}

\runninghead{Integrable $su(2)$-invariant spin chains}
{Integrable $su(2)$-invariant spin chains}

\begin{document}
\normalsize\textlineskip
{\thispagestyle{empty}
\setcounter{page}{1}

\renewcommand{\thefootnote}{\fnsymbol{footnote}} 

\copyrightheading{Vol. 0, No. 0 (1994) 000--000}

\vspace*{0.88truein}

\fpage{1}
\centerline{\bf INTEGRABLE $SU(2)$-INVARIANT SPIN CHAINS}
\vspace*{0.035truein}
\centerline{\bf AND THE HALDANE CONJECTURE}
\fnt{}{Paper submitted for the Proceedings of the
{\it Confronting the Infinite} Conference in honour of H. S. Green
and C. A. Hurst.}
\vspace{0.37truein}
\centerline{\footnotesize M. T. BATCHELOR and C. M. YUNG}
\vspace*{0.015truein}
\centerline{\footnotesize\it Department of Mathematics, Australian
National University}
\baselineskip=10pt
\centerline{\footnotesize\it Canberra, ACT 0200, Australia}
\vspace{0.225truein}

\vspace*{0.21truein}
\abstracts{\noindent
We perform a systematic exact algebraic search for integrable spin-$S$
chains which are isotropic in spin space, i.e. are $su(2)$-invariant.
The families of spin chains found for $S\le13.5$ support recent
arguments in favour of the complete classification of all such
integrable chains. The integrable families of spin chains are discussed
in the light of the conjectured spin-dependent properties of the
Heisenberg chain.
}{}{}

\vspace*{-3pt}\textlineskip
\section{Introduction}
\noindent
We are all familiar with the great interest of Green and Hurst
in exact solutions.
As emphasized in their beautiful book, such solutions
``have an intrinsic interest quite apart from the
illumination they impart to physical problems".\cite{gh}
Indeed, the study of exactly solved models has since evolved
into one of the key areas of mathematical physics, with the original
physical motivations left far behind in the wake.
In this paper, we also follow the original spirit of Green and Hurst
and pay attention to a particular physical phenomena.
In particular, we address the question of whether or not exactly
solved models have a bearing on the so-called
Haldane conjecture.\cite{haldane,aff}
This involves a surprise from quantum mechanics in one dimensional
systems.

Haldane argued that the properties of the Heisenberg antiferromagnet
\begin{equation}
 {\cal H} = \sum_{j=1}^{N}  \bm{S}_j\cdot \bm{S}_{j+1}
\end{equation}
should differ substantially if the spin $S$ is integer or half-odd integer.
For integer spin, there is a gap towards spin excitations,
whereas for half-odd integer spin, the model is massless with no gap.
Of course, for $S = \smallfrac{1}{2}$ the model was solved
long ago via the Bethe Ansatz\cite{bethe}, and subsequent investigations
revealed that the $S = \smallfrac{1}{2}$ model is indeed massless\cite{bbr}.
Evidence for the Haldane conjecture has been obtained
mainly for $S=1$ via explicit
numerical computations where the gap is of magnitude 0.41049(2),\cite{gapnum}
and ingenious experiments on related compounds,
in particular with NENP\cite{exp}.

In this paper we take up the systematic search for integrable isotropic
spin-$S$ chains with nearest-neighbour interaction.
Here our notion of an integrable Hamiltonian is one associated with an
$R$-matrix satisfying the Yang-Baxter equation.
There are of course very interesting spin chains which fall outside
this notion of integrability, such as the $S=1$ Hamiltonian
\begin{equation}
 {\cal H} = \sum_{j=1}^{N}  \bm{S}_j\cdot \bm{S}_{j+1} + \smallfrac{1}{3}
 \left( \bm{S}_j\cdot \bm{S}_{j+1}\right)^2,
\end{equation}
which possesses an exact valence bond groundstate and a gap\cite{aklt},
thus fulfilling the Haldane scenario. On the other hand, there are
three known integrable $S=1$ chains.
However, two of these chains have no gap. They are
\begin{equation}
 {\cal H} = \sum_{j=1}^{N}  \bm{S}_j\cdot \bm{S}_{j+1} +
 \left( \bm{S}_j\cdot \bm{S}_{j+1}\right)^2,
\end{equation}
discussed first by Uimin\cite{u}, and
\begin{equation}
 {\cal H} = \sum_{j=1}^{N}  \bm{S}_j\cdot \bm{S}_{j+1} -
 \left( \bm{S}_j\cdot \bm{S}_{j+1}\right)^2,
\end{equation}
which originates from the work of Kulish and Sklyanin\cite{ks}.
The remaining chain,
\begin{equation}
 {\cal H} =  - \sum_{j=1}^{N}  \left( \bm{S}_j\cdot \bm{S}_{j+1}\right)^2,
\end{equation}
does possess a gap, of magnitude $0.173\,178\ldots$
This is the biquadratic chain, discussed by a number of
authors\cite{p,bb,kl}.

The natural question arises, are there any other integrable $S=1$ chains
of this type? And more generally, what is known for arbitrary $S$?
In a recent paper Kennedy initiated a systematic search for such integrable
$su(2)$-invariant chains\cite{k}. A numerical
search for solutions of the Yang-Baxter equation for $S \le 6$
revealed four spin-$S$ families of integrable chains along with an
additional integrable chain at $S=3$.
More recently, we identified these $su(2)$-invariant chains with known
$\cal{G}$-invariant $R$-matrices\cite{O86,O87}, where ${\cal G}$
is a simple Lie algebra, and gave arguments that Kennedy's
results may well constitute the complete classification of such integrable
chains\cite{by}. These results are briefly reviewed
in the next section. In section 3 we extend Kennedy's search to $S \le 13.5$
by means of exact algebraic computation,
thus avoiding the possibility of missing any solutions due to roundoff error.
To conclude we discuss the integrable families
of spin chains in the light of the Haldane conjecture.

\newpage
\section{List of $su(2)$-invariant $R$-matrices}
\noindent
Integrable spin chains follow from the ``Master Key to Integrability"
-- the Yang-Baxter equation -- which we write
in the form (for reviews, see, e.g.\cite{ks,pa,j})
\begin{equation}
\left( \check R (\lambda) \otimes 1 \right) \left( 1 \otimes \check R
(\lambda + \mu) \right) \left( \check R (\mu) \otimes 1 \right) =
\left( 1 \otimes \check R (\mu) \right) \left( \check R (\lambda + \mu)
\right) \left( 1 \otimes \check R (\lambda) \right),
\label{ybe}
\end{equation}
with $\check R (0) = 1$. Given a solution $\check R (\lambda)$, the
Hamiltonian $H$ follows via the expansion
\begin{equation}
\check R (\lambda) = 1 + \lambda H + \sum_{n=2}^{\infty} \lambda^n
\check R^{(n)}.
\label{exp}
\end{equation}
It is thus clear that the search for $su(2)$-invariant Hamiltonians
is equivalent to the search for $su(2)$-invariant $R$-matrices.

Let ${\cal G}$ be a simple Lie algebra
of rank $n$ with fundamental weights denoted by $\Lambda_1,\ldots,\Lambda_n$.
Furthermore, let $\pi_{\Lambda}$  be an irreducible representation (irrep)
of ${\cal G}$ with highest weight $\Lambda$ on the vector space $V_{\Lambda}$.
Then the $R$-matrix $\check{R}^{\Lambda,\Lambda}(u) \in {\rm End}\left(
V_{\Lambda}\otimes V_{\Lambda}\right)$ is said to be ${\cal G}$-invariant if
\begin{equation}
[\;\check{R}^{\Lambda,\Lambda}(u),\;\pi_{\Lambda}({\cal G}) \otimes 1 +
1 \otimes \pi_{\Lambda}({\cal G})\;]=0.
\end{equation}
For a given pair $({\cal G},\Lambda)$ the imposition of
such a condition sometimes (but not always\cite{Drinfeld85,Mackay91})
allows the
Yang-Baxter equation for $\check{R}^{\Lambda,\Lambda}(u)$ to be solved.
In particular, for any ${\cal G}$ (except $E_8$)
and $\Lambda$ corresponding to the
lowest dimensional irreps the solutions are known explicitly.

Such a ${\cal G}$-invariant
$R$-matrix turns out also to be $su(2)$-invariant with spin $S$ if
the space $V_\Lambda$ can be identified with a space
$V_{2S\Lambda_1}$ on which $su(2)$ is represented {\em irreducibly}.
We were unable to give a complete classification of all pairs
$({\cal G},\Lambda)$ such that this condition holds. However, an examination
of the tables of branching rules\cite{McKay81} for simple Lie algebras
revealed only the solutions\cite{by}
\begin{description}
\item{(i)} $(A_n=su(n+1),\Lambda_1)$ for $n\ge 1$,
\item{(ii)} $(B_n=so(2n+1),\Lambda_1)$ for $n\ge 3$,
\item{(iii)} $(C_n=sp(2n),\Lambda_1)$ for $n\ge 2$, and
\item{(iv)} $(G_2,\Lambda_2,\Lambda_2)$.
\end{description}
In particular, we note that the $R$-matrices associated with the fundamental
representations of $D_n$, $E_6$, $E_7$ and $F_4$ are not $su(2)$-invariant.

We will now list the known $su(2)$-invariant $R$-matrices. In addition to
the spin $S$, we need extra labels $\{\rm I,IIa,IIb,III,IV\}$  to distinguish
between different families. In spectral form ($P^{(j)}$ are
projection operators onto $su(2)$-irreps
in $V_{2S\Lambda_1}\otimes V_{2S\Lambda_1}$), they are given by
\begin{eqnarray}
\check{R}^{2S\Lambda_1,2S\Lambda_1}_{\rm I}(u) & = & (1-u)\sum_{i\;{\rm
even}}P^{(i)}
  + (1+u)\sum_{i\;{\rm odd}}P^{(i)} \label{eqn:r1} \\
\check{R}^{2S\Lambda_1,2S\Lambda_1}_{\rm IIa} (u) & = &
   \left(1-u\right)\left(1-(S-\smallfrac{1}{2})u\right)P^{(0)} +
   \left(1+u\right)\left(1-(S-\smallfrac{1}{2})u\right)\sum_{i\;{\rm odd}}
   P^{(i)} + \nonumber\\
   & & \left(1+u\right)\left(1+(S-\smallfrac{1}{2})u\right)
   \sum_{i\;{\rm even}\; \ne 0} P^{(i)} \hspace{15pt} (S\; {\rm integer})
   \label{eqn:r2a} \\
\check{R}^{2S\Lambda_1,2S\Lambda_1}_{\rm IIb} (u) & = & \left(1-u\right)
   \left(1+(S+\smallfrac{3}{2})u\right)
   P^{(0)} + \left(1+u\right)\left(1+(S+\smallfrac{3}{2})u\right)
   \sum_{i\;{\rm odd}} P^{(i)} + \nonumber\\
  & & \left(1+u\right)\left(1-(S+\smallfrac{3}{2})u\right)
   \sum_{i\;{\rm even}\; \ne 0} P^{(i)} \hspace{15pt} (S\;
   {\rm half~odd~integer}) \label{eqn:r2b}\\
\check{R}^{2S\Lambda_1,2S\Lambda_1}_{\rm III}(u) & = & \sum_{k=0}^{2S}\left(
   \prod_{j=1}^{k}(j-u)\prod_{j=k+1}^{2S}(j+u)\right)P^{(k)}\\
\check{R}^{2S\Lambda_1,2S\Lambda_1}_{\rm IV}(u) & = & 1+ \frac{a-a e^u}
   {e^u-a^2}(2S+1) P^{(0)};
   \hspace{15pt} a+\smallfrac{1}{a} = 2S+1 \hspace{15pt} (S \ge 1).
\label{eqn:r5}
\end{eqnarray}
Here, it is understood that $\sum_{i\;{\rm even}}$ is short for
$\sum_{i=0\;(i\;{\rm even})}^{2S}$ etc.\
The extra solution for $S=3$ can be written as
\begin{eqnarray}
\check{R}^{6\Lambda_1,6\Lambda_1}_{\rm V}(u) & = &
   (1+6u)(1+u)(1-\smallfrac{3}{2}u) P^{(0)} +
   (1-6u)(1-u)(1+\smallfrac{3}{2}u) P^{(3)} + \nonumber\\
   & & (1-6u)(1-u)(1-\smallfrac{3}{2}u) \left(P^{(2)}+P^{(4)}+P^{(6)}\right)
   +\nonumber\\
   & & (1+6u)(1-u)(1-\smallfrac{3}{2}u) \left(P^{(1)}+P^{(5)}\right).
\label{eqn:s3}
\end{eqnarray}

The identification of these $su(2)$-invariant $R$-matrices
with $R$-matrices invariant under a larger algebra ${\cal G}$
is done with the help of the tabulated branching rules\cite{McKay81}.
The results are found to be\cite{by}
\begin{eqnarray}
\check{R}^{2S\Lambda_1,2S\Lambda_1}_{\rm I}\;\; \left[\;su(2)\;\right] & = &
  \check{R}^{\Lambda_1,\Lambda_1}\;\;[\;su(2S+1)\;]
  \hspace{15pt}( S \; {\rm half\; integer} ),\nonumber\\
\check{R}^{2S\Lambda_1,2S\Lambda_1}_{\rm IIa}\;\;[\;su(2)\;] & = &
  \check{R}^{\Lambda_1,\Lambda_1}\;\;[\;so(2S+1)\;]
  \hspace{15pt}( S \;{\rm integer} ),\nonumber\\
\check{R}^{2S\Lambda_1,2S\Lambda_1}_{\rm IIb}\;\;[\;su(2)\;]& = &
  \check{R}^{\Lambda_1,\Lambda_1}\;\;[\;sp(2S+1)\;]
  \hspace{15pt}( S \;{\rm half\; odd\; integer} ),\nonumber\\
\check{R}^{6\Lambda_1,6\Lambda_1}_{\rm V}\;\;[\;su(2)\;] & = &
  \check{R}^{\Lambda_2,\Lambda_2}\;\;[\;G_2\;]
\qquad\qquad \;\;\; (S = 3)\nonumber.
\end{eqnarray}
The $R$-matrices $\check{R}^{2S\Lambda_1,2S\Lambda_1}_{\rm III}[su(2)]$
correspond to the trivial embedding of $A_1$ in itself and are thus
already in the ``proper'' Lie algebraic setting. The remaining solution,
$\check{R}^{2S\Lambda_1,2S\Lambda_1}_{\rm IV}[su(2)]$ is not rational
in $u$, unlike the others under consideration, and is in a class of
its own --  being related to the Temperley-Lieb algebra.

\section{Systematic Exact Search}
\noindent
In order to lend further weight to the above list of
$su(2)$-invariant $R$-matrices and hence integrable $su(2)$-invariant
spin chains being complete, we turn now to an exact systematic search of
solutions of the Yang-Baxter equation following Kennedy\cite{k}.
Our interest lies in quantum chains with Hamiltonians
\begin{equation}
{\cal H} = \sum_{j=1}^{N} H_{j,j+1}
\end{equation}
such that $H_{j,j+1}$ is a copy of $H$ acting on sites $j$
and $j+1$, which in turn is $su(2)$-invariant. As in the above, we restrict our
attention to models for which the spin $S$ at each site is the same.

The condition of $su(2)$-invariance implies that $H$ can be written as a linear
combination of the $su(2)$-projectors $P^{(j)}$ for $j=0,\ldots,2S$.
As we are interested in the Hamiltonian only up to constants (and since
$1 = \sum_{j=0}^{2 S} P^{(j)}$ is the resolution of the identity), it is
sufficient to set
\begin{equation}
H = \sum_{j=0}^{2 S - 1} c_j P^{(j)},
\end{equation}
where the coefficients $c_j$ are to be determined. The projectors can be
explicitly constructed from the matrix representatives of the Casimir
operator. For large $S$, it is more efficient to construct them from
Clebsch-Gordan coefficients.

Given the expansion (\ref{exp}), a
necessary condition for $\check{R}(\lambda)$ to satisfy the
Yang-Baxter equation (\ref{ybe}) is Reshetikhin's condition\cite{ks,k}
\begin{equation}
\left[ H \otimes 1 + 1 \otimes H, \left[ H \otimes 1, 1 \otimes H \right]
\right] = 1 \otimes X - X \otimes 1,
\end{equation}
for some operator $X$. Following Kennedy\cite{k}, we can simplify this
condition further for $su(2)$-invariant chains by taking the matrix elements
$\langle S, j, k | \cdots | j + k - S, S, S \rangle$ ($k<S$ and $j+k\ge0$),
which results in the right hand side vanishing and hence $X$ drops out of
consideration. We note that this simplified Reshetikhin condition is
necessary but not sufficient for the Reshetikhin condition to hold, which
itself is necessary but (possibly) not sufficient for the YBE to hold.

When the simplified Reshetikhin condition is expanded out, we obtain
\begin{equation}
\langle S, j, k | (1 \otimes H) (H^2 \otimes 1) - 2 (H \otimes 1)(1 \otimes H)
(H \otimes 1) - \cdots |j + k - S, S, S \rangle = 0,
\end{equation}
for $k < S$ and $j+k\ge0$. This can be evaluated in terms of the matrix
elements of $H$ which in turn are given by
the known projector matrix elements and the unknown constants $c_j$.
Reshetikhin's condition (simplified) then reduces to a system of homogeneous
cubic equations in the $c_k$'s, which can be solved recursively. At each
step of the recursion only a quadratic equation needs to be solved.

We have solved this system of equations algebraically using {\em Mathematica}
for $S \le 13.5$, which is a significant extension of the numerical
calculations\cite{k} performed for $S\le 6$. We have confirmed that
no new solutions appear. The only integrable $su(2)$-invariant spin chains
we see for $S \le 13.5$ are those following from the $R$-matrices listed in
(9)-(14).

\section{List of $su(2)$-invariant Spin Chains}
\noindent
The families of integrable spin chains under discussion are all of the
form (15), (16). The first family has two-body interactions
\begin{equation}
H_{\rm I} = {\cal P},
\end{equation}
where ${\cal P}=(-)^{2S}\sum_{i=0}^{2S}(-)^i P^{(i)}$ is the permutation
or exchange operator. The second family
can be written in the combined form\cite{k}
\begin{equation}
H_{\rm II} = \left[ S + \smallfrac{1}{2} - (-)^{2S} \right] {\cal P}
 - (-)^{2S} (2S+1) P^{(0)}.
\end{equation}
For the third family,
\begin{equation}
H_{\rm III} = \sum_{k=0}^{2S} \left( \sum_{j=1}^k \frac{1}{j} \right) P^{(k)}.
\end{equation}
The Temperley-Lieb family is simply
\begin{equation}
H_{\rm IV} = P^{(0)}.
\end{equation}
The $S=3$ $G_2$ chain has Hamiltonian
\begin{equation}
H_{\rm V} = 11 P^{(0)}+7 P^{(1)}-17 P^{(2)} -11 P^{(3)} -17 P^{(4)}
+ 7 P^{(5)} -17 P^{(6)}.
\end{equation}
All of the above expressions follow from the $su(2)$-invariant $R$-matrices
of the preceding section up to overall multiplicative and additive factors.
They can all be alternatively written, subject to the same caveat, in terms
of the more familiar spin operators $X_j =\bm{S}_j\cdot \bm{S}_{j+1}$ via
\begin{equation}
P^{(i)} = \prod_{\stackrel{k=0}{\ne i}}^{2 S}
{X_j - x_k \over x_i - x_k},
\end{equation}
where $x_k = \smallfrac{1}{2} k (k+1) - S(S+1)$. In this way we see,
for example, for $S=1$, $H_{\rm I}$ reduces to the Hamiltonian (3), while
$H_{\rm II}$ and $H_{\rm III}$ reduce to (4) and $H_{\rm IV}$ reduces to (5).

\section{Concluding Remarks}
\noindent
We first remark that there are no other integrable $su(2)$-invariant
$S=1$ chains beyond the three mentioned in the Introduction.
Apart from the fifth solution at $S=3$, there are four integrable
spin chains for each value of $S$ when $S>1$.
We believe that this latter statement holds for arbitrary $S$.

Among the integrable $su(2)$-invariant spin chains
we do not see the Heisenberg chain (1). One may harbour
a faint hope that it may appear for some high value of $S$ by some
miraculous vanishing of coefficients in one of the integrable families.
However, we have not seen this to be the case for $S$ up to 100.
Thus it appears most unlikely that, apart from the $S=\smallfrac{1}{2}$ case,
an integrable spin chain can yield either an exact confirmation or
counterexample to the Haldane conjecture. Nevertheless, it is still
of interest to investigate the properties
of the integrable chains in light of the conjectured properties of
the Heisenberg chain.

Among the integrable families of spin chains, the Temperley-Lieb chains
appear to be the only ones with a gap.
This gap is known exactly and opens up with increasing $S$.\cite{bbs,kls}
However, the gap exists for all $S\ge 1$. So in this case
we see no dramatically different behaviour depending on whether
$S$ is integer or half-odd integer. The remaining families of spin chains
all appear to be critical for all $S$, with no gap.
Nevertheless, there does appear to be an
integer {\it vs} half-odd integer distinction in the Hamiltonian (20) arising
from family II. Recall that this chain is associated with
$so(2S+1)$ for $S$ integer, and with $sp(2S+1)$ for $S$ half-odd integer.
For critical models, it is now both rather well understood and confirmed
for a number of cases\cite{dvl} that the central charge defining the
underlying universality class is given by
\begin{equation}
c = {m\, {\cal D} \over m + h},
\end{equation}
where ${\cal D} = {\rm dim}\, {\cal G}$, $h$ is the Coxeter number and
$m$ is the level of the representation.
Thus for ${\cal G}=B_n=so(2n+1)$, where ${\cal D}=n(2n+1)$, $h=2n-1$,
$m = 1$ and $n=S$, we expect the value $c = S+\smallfrac{1}{2}$.
On the other hand, for ${\cal G}=C_n=sp(2n+1)$, where ${\cal D}=n(2n+1)$,
$h=n+1$, $m = 1$ and
$2n=2S+1$, we expect the value $c = (2S+1)(2S+2)/(2S+5)$.
Work to confirm these values is currently in progress. In particular,
for the $so(2S+1)$ family, the Bethe Ansatz roots defining the groundstate
form a sea of 2-strings in the complex plane for all $S$,\cite{dvl} thus
allowing a direct application of a recently developed method for deriving
the central charge for such cases via the dominant finite-size correction
to the groundstate energy\cite{kb,kbp,wbn}.

\nonumsection{Acknowledgements}
\noindent
It is a great pleasure for us to wish Professors Green and Hurst
all the best for their 8th decade.
We thank T. Kennedy for a helpful correspondence.
This work has been supported by the Australian Research Council.

\nonumsection{References}

\end{document}